\documentclass[letter]{ieeeconf}
\IEEEoverridecommandlockouts
\usepackage[utf8]{inputenc}
\usepackage[margin=1in]{geometry}
\usepackage{amsmath,amssymb,amsfonts}
\usepackage{bm}% bold math
\usepackage{color}
\usepackage{verbatim}
\usepackage{lmodern}
\usepackage{parskip}
\usepackage[pdftex]{graphicx}
\usepackage{epsfig}
\usepackage{psfrag}
\usepackage{tabls}
\usepackage{caption}
\usepackage{dsfont}
\usepackage{tikz}
\usepackage{framed}
\usetikzlibrary{automata,positioning}
\usepackage{times}
\usepackage{amsfonts,amscd}
\usepackage{multicol}
\usepackage{caption}
\usepackage[utf8]{inputenc}
\usepackage{lscape}
\let\labelindent\relax
\usepackage{enumitem}
\usepackage{epstopdf}
\usepackage[ruled,vlined]{algorithm2e}
\usepackage{hyperref}
\usepackage{mathtools}
\usepackage{dsfont}
\usepackage{bibentry}
\nobibliography*
\usepackage{hyperref}
\newcommand{\ip}[2]{\left\langle #1 , #2 \right\rangle}
\newcommand{\ddt}{\frac{\mathrm{d}}{\mathrm{d}t}}
\newcommand{\bp}[1]{\Big( #1 \Big)^\top}

\newcommand{\nn}{\nonumber}
\newcommand{\e}{\boldsymbol{e}_3}
\newcommand{\bR}{\mathbb{R}}
\newcommand{\bI}{\mathbb{I}}

\newcommand{\hatsq}[1]{{(\hat{#1})}^2}
\newcommand{\norm}[1]{{\left \lVert{#1}\right \rVert}}

\newcommand{\deff}{\coloneqq}
\newcommand{\rta}{\Rightarrow}

\newtheorem{assumption}{Assumption}

\newtheorem{lemma}{Lemma}[section]
\newtheorem{definition}{Definition}[section]
\newtheorem{remark}{Remark}

\newtheorem{theorem}{Theorem}

\title{Stabilizing a spherical pendulum on a quadrotor}

\author{Aradhana Nayak, Ravi N Banavar and D. H. S. Maithripala
\thanks{U{\tt\small bb}}%
}
\begin{document}
\maketitle
\thispagestyle{empty}
\pagestyle{empty}
% These force using more of the margins that is the default style
% This command causes the title to be created in the document
\begin{abstract}
In this article we design a backstepping control law based on geometric principles to swing up a spherical pendulum mounted on a moving quadrotor. The available degrees of
freedom in the control vector also permit us to position the plane of the
quadrotor parallel to the ground. The problem addressed here is, indeed,  novel and has many practical applications which arise during the transport of a payload mounted on top of a quadrotor. The modeling and control law are coordinate-free and thus avoid singularity issues. The geometric treatment of the problem greatly simplifies both the modeling and control law for the system. The control action is verified and supported by numerical experiments for aggressive manoeuvres starting very close to the downward stable equilibrium position of the pendulum.
\end{abstract}

\section{Introduction}
%
%\textcolor{blue}{Introduction is not polished}
%
The problem of achieving an arbitrary orientation and tracking of suitable trajectory for a quadrotor is well studied in literature (\cite{tleemleok}, \cite{lee2010geometric}). A quadrotor consists of four arms with rotors attached to them. The propeller attached to the rotors thus provide four independent directions of actuation. $3$ of these can be utilized to achieve an arbitrary position in $\bR^3$. In \cite{lee2010geometric}, the fourth actuation is used to track a heading direction of the quad. Most commercially available UAVs can be modelled as quadrotors. The availability of sophisticated and affordable sensors in the recent years has led to large scale manufacturing of UAVs. This has led to their utilization in transporting load over reasonably long distances. The load is usually considered to be suspended by a cable attached to the center of the quadrotor. Cable suspended systems are underactuated and therefore, there has been an increased effort in the robotics community to study the various control objectives which can be realised by such systems.

In this article, we aim to balance an inverted pendulum mounted on the center of mass of a quad through a universal joint. This mechanical system, called flying inverted pendulum was first introduced in \cite{dandrea}.  The flying pendulum is the simplest model for a payload mounted on a quadrotor. Therefore the stability of such a mechanical system is a potentially important problem which has not been addressed in the literature. The problem of a payload suspended through cables from multiple quadrotors has recently gained a lot of attention and is well studied in \cite{lee2013geometric}, \cite{lee2018geometric} and \cite{wu2014geometric}. In such a payload mounting, there can be issues of damage to the payload during landing of the quad. In the proposed model, a more practical method of transporting the payload is achieved by mounting it on top of the quad.

A linearization approach was used in \cite{dandrea} to stabilize the pendulum on the quadrotor. Specifically, nominal trajectories of the system were considered and the system was stabilized around these dynamical equilibria. This is a fairly restrictive treatment as the nonlinearities in the quadrotor can lead to failure of the control action if the initial position of the pendulum is far away from the equilibria, for example for an initial condition close to the inverted equilibrium position.
Herein lies the novelty of our contribution. In this article, the quadrotor and pendulum are modelled as Lagrangian systems and the control law proposed admits convergence to the desired state from
a large set of initial conditions, significantly away from the desired orientation, and thereby allowing aggressive manoeuvres. To make it as realistic to a real-life situation, the parameters of
the pendulum-quadrotor assembly considered in this paper matches that of the experimental set-up in \cite{dandrea}. The salient advantage of the control action, apart from being nonlinear and continuous, is that is coordinate-free. In \cite{dandrea}, the orientation of the quad is modelled using Euler angles which suffer from singularity issues and hence do not allow the control action to achieve aggressive manoeuvres. In this article, however, we employ a purely geometric model: the orientation of the quad is modelled as a rotation matrix and the pendulum is modelled as a point mass on the $2-$ sphere, thereby accounting for all possible configurations which the system may assume.

The contributions of this article are: 1. Explicit control laws in continuous time for the swing up of an inverted pendulum mounted using a spherical joint on a quadrotor, 2. Geometric modelling for the study of dynamics and control of the system which allows a large set of initial conditions from which stability is guaranteed. 3. Strict feedback form of the part of the dynamics which has to be controlled is shown to exist which allows backstepping control to be applied. 4. Backstepping control is used in a purely geometric setting by choosing appropriate Lyapunov functions.

The article is organized as follows: In Section II we inbtroduce the notations and the dynamical model, in Section III we propose the control strategy by establishing the strict feedback form of the dynamical equations, in Section IV we present numerical experiments with model parameters chosen close to an experimental set-up by choosing several non trivial initial conditions, and finally we conclude the paper in Section V.

\section{Notation and Dynamic Modelling}
\subsection{Notation}
\begin{figure}[h!]
  \centering
  \includegraphics[scale=0.5]{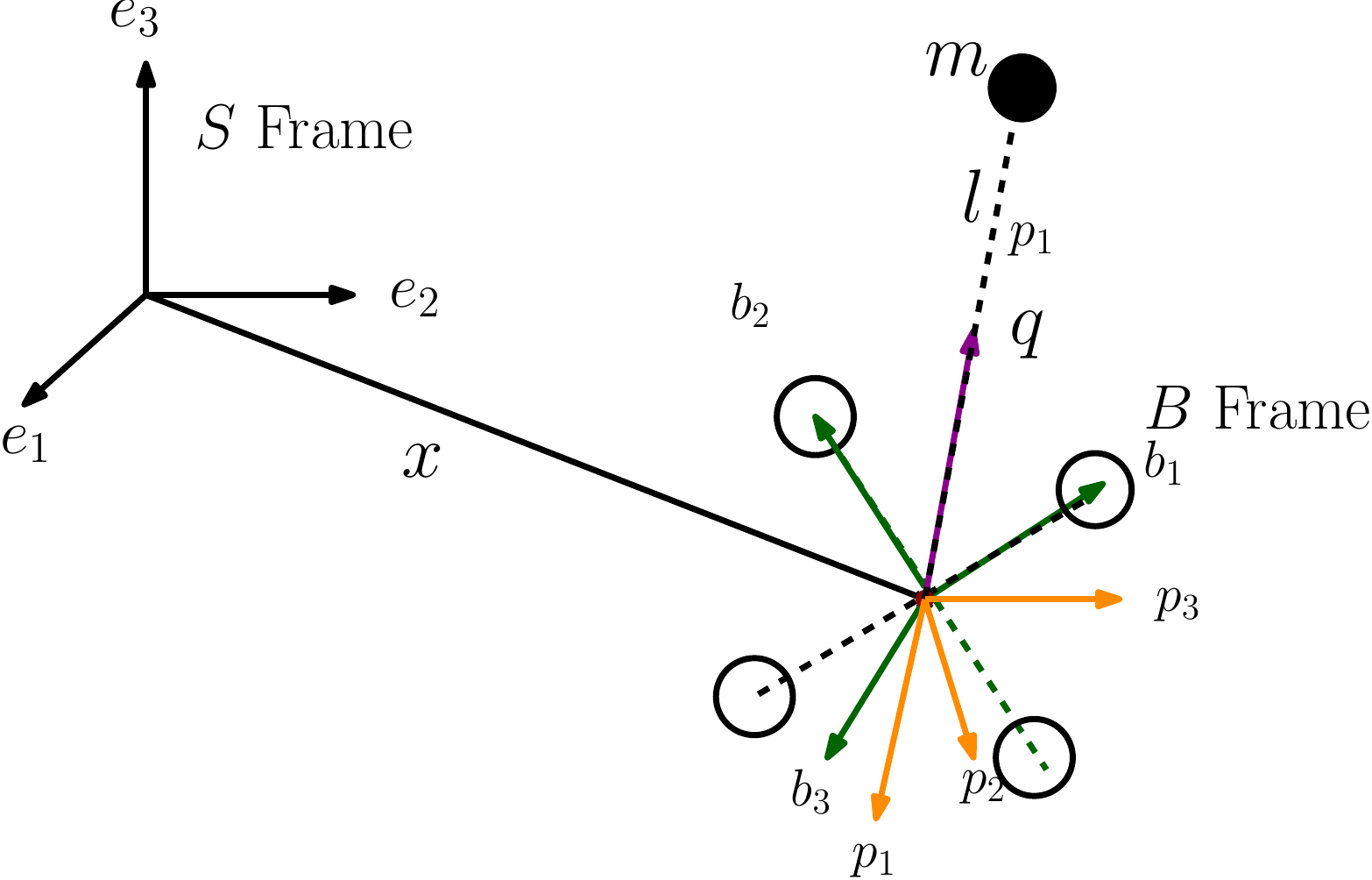}
  \caption{Pendulum on Quadrotor}\label{diag}
\end{figure}

Figure \ref{diag} represents the quadrotor with a pendulum mounted on it.
The following notation is employed:
%
%\textcolor{magenta}{May I suggest using better notation so that understanding improves later on ?
%Fix two frames - one on the pendulum ($b_1,b_2,b_3$ with $b_3$ along the pendulum link,
%and one on the quad - $q_1,q_2,q_3$. ) Let the rotation of $b$ with respect to $q$ be $R_p$,
%and that of $q$ with respect to the spatial frame be $R_q$. }
%
%\bei
%
\begin{itemize}
\item $S$: The spatial frame specified by vectors $\{ \boldsymbol{e}_1 ,\boldsymbol{e}_2,\e\}$ where $\boldsymbol{e}_1= \begin{pmatrix}
                  1 & 0 & 0
                \end{pmatrix}^\top$, $\boldsymbol{e}_2= \begin{pmatrix}
                  0 & 1 & 0
                \end{pmatrix}^\top$, $\e= \begin{pmatrix}
                0 & 0 & 1
                \end{pmatrix}^\top$
	\item $B$: The frame fixed to the center of mass of the quadrotor (pivot) specified by vectors $\{ b_1,b_2,b_3\}$
    \item $P$: The frame fixed to the pivot specified by vectors $\{ p_1,p_2,p_3\}$
	\item $m$ and $M$: Mass of the pendulum and quadrotor respectively
	%\item $\vec{x}_m$: position vector of the pivot of the pendulum expressed in the spatial frame
\item $\bI$: Moment of inertia of the quadrotor in $B$ frame
	\item $l$: Length of the inextensible cable connecting the center of mass of the quadrotor (pivot) to the mass $m$
    \item $d$: Distance from the pivot to the rotors
	\item $x$: Location of the pivot in the $S$ frame
	\item $q$: Unit vector from the pivot along $-p_1$ in the $B$ frame
	\item $R$: Rotation matrix expressing the transformation from the $B$ frame to the $S$ frame, $R b_i = e_i$.
	\item $y$:  Unit vector from the pivot along the pendulum in the $S$ frame, $y = Rq$.
	\item $\omega$: Angular velocity of the quadrotor expressed in the quadrotor body in $B$ frame
	\item $g$: Gravity vector in the $-\e$ direction
    \item $f_i$: Magnitude of thrust generated by $i$-th propeller along $-b_3$
    \item $f$: Magnitude of total thrust, $f= \sum_{i=1}^{4} f_i $
    \item $\tau_i$: Torque generated by the $i$-th propeller about the $b_3$ axis
\end{itemize}
%
%\ei
%
All the $4$ rotors are in the $b_1-b_2$ plane. We consider a universal (spherical) joint at the pivot and the bob of the pendulum to be symmetric. Therefore, $q$ completely specifies the $P$ frame as we ignore the rotation about $p_1$ axis. By definition, the total thrust is $-f R \e$ in the inertial frame. It is assumed that the first and the third propellers rotate clockwise, and the second and the fourth propellers rotate counterclockwise, when they are generating a positive thrust $f_i$, the torque generated by the $i$-th propeller can be written as $\tau_i = {(-1)}^i c f_i$ for a fixed constant c. Under these fairly common assumptions (see \cite{lee2018geometric} and \cite{lee2010geometric}) the total thrust $f$ and the total moment $\mu = \begin{pmatrix}
  \mu_1 & \mu_2 & \mu_3
\end{pmatrix}^\top$ acting at the pivot can be transformed as
\begin{equation}\label{transfmu}
  \begin{pmatrix}
    f \\
    \mu_1 \\
    \mu_2 \\
    \mu_3
  \end{pmatrix} = \begin{pmatrix}
                    1 & 1 & 1 & 1 \\
                    0 & -d & 0 & d \\
                    d & 0 & -d & 0 \\
                    -c & c & -c & c
                  \end{pmatrix} \begin{pmatrix}
                                  f_1 \\
                                  f_2 \\
                                  f_3 \\
                                  f_4
                                \end{pmatrix}.
\end{equation}
Since the transformation is invertible, in this article, $\begin{pmatrix}
    f &    \mu
  \end{pmatrix}^\top $ is considered as the control input to the quadrotor-pendulum system. The dynamical model considered in this paper is hence based on the experimental set-up in \cite{dandrea}. We do not assume the mass of the pendulum is small apriori, however the rigorous modelling allows us to conclude mathematically that the quadrotor dynamics is not affected by that of the pendulum as assumed in \cite{dandrea}.  All control algorithms we propose can therefore be applied to an experimental quadrotor-pendulum assembly directly.

The kinetic energy of the quadrotor is  denoted as $L_Q$, and that of pendulum is denoted as $L_p$. The potential energy is denoted by $V$. The total lagrangian is
\begin{equation}\label{L}
   L= L_Q + L_p -V
\end{equation}
wherein the individual expressions are as follows:
\begin{equation}\label{lq}
    L_Q = \frac{1}{2} M \ip{\dot{x}}{\dot{x}} + \frac{1}{2}  \ip{\mathbb I \omega}{\omega},
\end{equation}
\begin{equation}\label{lp}
    L_p = \frac{m}{2}\ip{(\dot{x} + l\dot{y})}{(\dot{x} + l\dot{y})},
\end{equation}
and
\begin{equation}\label{pe}
    V= -\ip{Mg x}{\e} - \ip{mg(x+ly)}{ \e}.
\end{equation}
 The variation $\delta L$ of the Lagrangian is given by:
\begin{align*}
  \delta L &=  \ip{\frac{\partial L}{\partial x}}{\delta x }
    +\ip{\frac{\partial L}{\partial \dot{x}}}{\delta \dot{x}}
    + \ip{\frac{\partial L}{\partial \omega}}{\delta \omega } \\
   &+\ip{\frac{\partial L}{\partial y}}{\delta y }
    + \ip{\frac{\partial L}{\partial \dot{y}}}{\delta \dot{y}}
\end{align*}
The variations of the individual terms are:
\begin{alignat*}{7}
   &\text{In }\dot{x}: &\delta \dot{x}= \frac{\mathrm{d}}{\mathrm{d}t}(\delta x),\\
  &\text{In }R :  &\hat{\Sigma}\deff R^\top \delta R,\\
   &\quad &\dot{\hat{\Sigma}}= - R^\top \delta R R^\top \dot{R} + R^\top \delta \dot{R}, \\
  &\text{In }\omega: &\quad \delta \hat{\omega} = - R^\top \delta R R^\top \dot{R} + R^\top \delta \dot{R} = [\hat{\omega}, \hat{\Sigma}] + \dot{\hat{\Sigma}}, \\
 &\text{In } y:    &\delta y = \delta R q + R \delta q = R\hat{\Sigma} R^\top y + \gamma \times y, &\quad\\
  &\text{In } \dot{y}:  & \delta\dot{y} = \ddt{} (R\hat{\Sigma} R^\top y) +  \dot{\gamma} \times y + \gamma \times \dot{y} &\quad\\
  &\quad &= R ([\hat{\omega}, \hat{\Sigma}] + \dot{\hat{\Sigma}})R^\top y +  R\hat{\Sigma} R^\top \dot{y} + \dot{\gamma} \times y + \gamma \times \dot{y}\\
   &\quad &\text{for any }\gamma \in \mathbb{R}^3 \text{s.t.} \gamma^ \top y= 0.
\end{alignat*}
\subsection{Dynamic Model}
In this section we elaborate on derivation of the dynamical equations of the of the quadrotor-pendulum assembly in Figure \ref{diag} by variational principles considering the lagrangian in $L$ in \eqref{L}. The method we employ is truly coordinate free as all configuration variables are considered to lie in an ambient Euclidean space and their corresponding variations are constrained by the geometry of the manifold on which the configuration variables evolve.

The partial derivatives of the Lagrangian $L$ are:
\begin{alignat*}{3}
&\bp {\frac{\partial L}{\partial x}} &=(M+m)g \e ; \bp{ \frac{\partial L}{\partial \dot{x}}} &= {(m+M)} \dot{x} + ml \dot{y}\\
  & \bp {\frac{\partial L}{\partial \hat{\omega}}} &=  \mathbb{I} \hat{\omega};  \bp {\frac{\partial L}{\partial y}} &= mgl \e ;\\
 & \bp {\frac{\partial L}{\partial \dot{y}} } &= ml \dot{x} +  ml^2 \dot{y}.
\end{alignat*}
According to the variational principle for nonconservative systems,
\begin{equation}\label{hamprin}
    \int_{0}^{T} \delta L (\Gamma, \dot{\Gamma})\mathrm{d}t=  \int_{0}^{T} F(\Gamma,\dot{\Gamma}) \delta\Gamma \mathrm{d}t
\end{equation}
 for curves $\Gamma:[0,T] \to \mathbb{R}^3 \times SO(3) \times \mathbb{R}^3$ with fixed end points: $\Gamma(0)= \Gamma_0$ and $\Gamma(T)= \Gamma_T$, and for generalized forces along the curves given by $F(\Gamma,\dot{\Gamma})$,  yields the equations of motion. The RHS of \eqref{hamprin} is the time integral of the virtual work over the interval $[0,T]$.  The details of this derivation are found in \cite{arxivpenduquad}.
 %

 %$We now incorporate the actuating forces into our model. Consider a force $f_i$, $i=1, \dotsc, 4$ acting on each rotor in $\e$ direction as shown in Figure \ref{diag}. The total moment acting on c.o.m of quadrotor is $\mu \in \bR^3$ and we denote the total thrust by $f \in \bR$ as $f= \Sigma_{i=1}^4 f_i$. It can be shown that there is an invertible transformation between the vectors $\pmat{f_1 \\ f_2 \\ f_3 \\ f_4}$ and $\pmat{f \\ \mu}$. Therefore, we consider the external force to be $\pmat{f \\ \mu}$. %

Denoting the vector $z \coloneqq R \e$, the equations of motion are:
 %
 %%%%%%%%%
 %
\begin{equation}\label{xeqn_txt}
    -(m+M) \ddot{x} - ml \ddot{y} + (M+m)g \e= -fz.
\end{equation}
\begin{align}\label{Rfin}
   \bI \dot{\omega} + ( \omega \times \bI  \omega ) = -\mu
    \end{align}
    \begin{equation}\label{ddty_notfin}
      \ddot{y} +\norm{\dot{y}}^2y =  \frac{f}{Ml} {(\hat{y})}^2 z
\end{equation}

\section{Control Strategy}
The objective is to choose $\mu$ and $f$ so as to stabilize the pendulum in the inverted upright position and position the quadrotor-plane  parallel to the ground. Mathematically
stated, the first and second requirement translate to
\begin{align*}
	\lim_{t \rta \infty} y(t) \rta \e  \;\;\;\; \lim_{t \rta \infty}  z(t) =  \lim_{t \rta \infty} R(t)  \e = \e
\end{align*}
From the definition of $z$ and differentiating twice we have
$   \ddot{z}= R( {\hat{\omega}}^2 + \dot{\hat{\omega}} )\e $
The dynamical equations for the purpose of
control design, with $\mu$ and $f$ being the control variables, are:
\begin{framed}
\begin{align*}
   \dot{R} = R \hat{\omega}, \quad \bI \dot{\omega} + ( \omega \times \bI  \omega ) = -\mu
    \end{align*}
\begin{equation}\label{ddtz}
    \ddot{z} = R( {\hat{\omega}}^2 +  \{\bI^{-1}(\bI \omega\times \omega)  -\bI^{-1} \mu  \}^{\hat{}})\e \quad \hbox{Quad Eqn}
%      \ddot{z} = R( {\hat{\omega}}^2 + \hat { \{\bI^{-1}(\bI \omega\times \omega)  -\bI^{-1} \mu  \} })\e
\end{equation}
    \begin{equation}\label{ddty}
      \ddot{y} +\norm{\dot{y}}^2y =  \frac{f}{Ml} {(\hat{y})}^2 z \quad \hbox{Pendulum Eqn}
\end{equation}
\end{framed}

\begin{assumption}
The controller has full access to the state at all times using appropriate sensors on board the quadrotor.
\end{assumption}
%
%\begin{assumption}
%%
%The initial conditions of the quadrotor and the pendulum are restricted in the following sense: the initial condition of the pendulum (in inertial frame) is away from the plane orthogonal to gravity vector and the plane of the quadrotor does not contain the gravity vector. In other words, $y(0) \neq - \e$ and $z(0) \neq -\e$.
%%
%\end{assumption}
%%
\subsection{Control on the $2$- sphere}
Both the subsystems to be controlled, namely the $z$ and the $y$ variables, evolve on a two-dimensional
sphere (or the $2$- sphere). We first present a few preliminaries of control on a sphere.

\begin{definition}
A fully actuated simple mechanical system (an SMS) on the $2$- sphere denoted by $S^2$ is specified by the $3$ tuple- $(S^2, I_3, u)$ where $I_3$ is the Euclidean metric on $S^2$ and $u \in \bR^3$ is the control vector. The equations for the controlled SMS $(S^2, I_3, u)$ are as follows.
\begin{equation}\label{SMSS2}
\nabla_{\dot{\phi}} \dot{\phi} \deff    \ddot{\phi}(t) + \norm{\dot{\phi}}^2 \phi(t) = -{(\hat{\phi})}^2u
\end{equation}
where $\phi(t) \in S^2$ is the controlled trajectory and $\nabla$ is the affine connection corresponding to the Euclidean metric.
\end{definition}
\begin{definition}\label{fullact}
The SMS $(S^2, I_3, u)$ is said to be fully actuated if the control forces $u$ generate the cotangent bundle $T^*S^2$.
\end{definition}
We now state two useful lemmas for asymptotic stabilization about a set point and asymptotic tracking of a reference trajectory for an SMS on $S^2$.
\begin{lemma}{Regulation on $S^2$}\label{lem1}
Consider the fully actuated SMS in \ref{SMSS2}. The following control law ensures that $\phi(t)$ is asymptotically stable about $\e \in S^2$
\begin{equation}\label{ureqd}
   u = -k_p \hat{\phi}^2\e -k_d \dot{\phi}
\end{equation}
%Furthermore, $\phi(t)$ converges to $\e$ exponentially if,
%\begin{equation}\label{initcon11}
%  k_p> \frac{\norm{\dot{\phi}(0)}^2}{2(1+ \phi(0)^\top \e)} .
%  \end{equation}
\end{lemma}

\proof
 Lemma 11.7 in \cite{bulolewis}
\endproof
\begin{lemma}{Tracking a trajectory on $S^2$}\label{lem2}
Consider the fully actuated SMS in \ref{SMSS2}. The following control law ensures that $\phi(t)$ asymptotically tracks a smooth and bounded reference trajectory $\phi_d(t) \in S^2$
\begin{equation}\label{ureqd1}
   u = -k_p \hat{\phi}^2\e -k_d v_e + \nabla_{\dot{\phi}} (\tau(\phi,\phi_d) \dot{\phi}_d)
\end{equation}
where, the velocity error $v_e \in T_{\phi} S^2$ is
\[  v_e \deff \dot{\phi} - \tau(\phi,\phi_d) \dot{\phi}_d ,\]
 the transport map $\tau(\phi,\phi_d): T_{\phi_d} S^2 \to T_{\phi} S^2$ is
 \[
 \tau(\phi,\phi_d) \dot{\phi}_d \deff (\phi_d \times \dot{\phi}_d) \times \phi,
 \]
and $\nabla_{\dot{\phi}} (\tau(\phi,\phi_d) \dot{\phi}_d)$ is the feedforward part of the control which simplifies as follows
\begin{equation*}
  \nabla_{\dot{\phi}} (\tau(\phi,\phi_d) \dot{\phi}_d) = \ip{\phi}{\phi_d \times \dot{\phi}_d}(\phi \times \dot{\phi}) + (\phi_d \times \ddot{\phi})\times \phi
\end{equation*}
%Furthermore, the convergence of $\phi(t)$ to the reference trajectory $\phi_d(t)$ is exponential if $k_p$, $\phi(0)$, $\dot{\phi}(0)$ , $\phi_d(0)$ and $\dot{\phi}_d(0)$ satisfy the following condition
%\begin{equation}\label{initcon21}
%  k_p> \frac{\norm{v_e(0)}^2}{2(1+ \phi(0)^\top \phi_d(0))}.
%  \end{equation}
\end{lemma}
\proof
 Section 11.3.2 in \cite{bulolewis}
\endproof

%\proofover
\subsection{Approach to controller design}
%\begin{equation}\label{ddty}
%     \ddot{y} + \norm{\dot{y}}^2 y =  \frac{f}{Ml} {(\hat{y})}^2 z
%\end{equation}
Feedback regularization (\cite{madhuACC}) refers to the use of feedback to impart the structure of an SMS to a fully actuated mechanical system.
Once a system is feedback regularized, a straightforward PD control action could then be employed
to stabilize the system. It is observed from \eqref{ddty} that the only control variable is $f$ which is one dimensional. The cotangent bundle of $S^2$ can be generated by at least $3$ independent covector fields (\cite{milnor1978analytic}). Therefore in order to attain an arbitrary configuration on $S^2$, at least $3$ independent directions of control are necessary. From Definition \ref{fullact} it means that the system described by equation \eqref{ddty} is not fully actuated. Therefore feedback regularization cannot be utilized to stabilize $y(t)$ about $\e$.

It is also observed that the output $z(t)$ affects the acceleration of the pendulum. Therefore, if we show that the set of equations \eqref{ddty} and \eqref{ddtz} are in strict-feedback form (\cite{khalil2002nonlinear}), a backstepping control can be used to stabilize $y(t)$ by choosing an appropriate $z(t)$ as an intermediate control for the equation \eqref{ddty}.
Since $\mu \in \bR^3$ appears in \eqref{ddtz}, therefore there are three independent control directions and \eqref{ddtz} is fully actuated. Therefore $z(t)$ can be reach an arbitrary configuration on $S^2$ using feedback regularization.

%\textcolor{blue}{We now present the procedure of control design}
%\\  \\
%
%\textcolor{magenta}{
\noindent {\it Philosophy of control design}:  Consider the two equations \eqref{ddtz}-\eqref{ddty}
 %
 %\begin{framed}
%%
%    \begin{equation*}
%      \ddot{y} +\norm{\dot{y}}^2y =  \frac{f}{Ml} {(\hat{y})}^2 z \quad \hbox{Pendulum equation}
%\end{equation*}
%%
%\begin{equation*}
%    \ddot{z} = R( {\hat{\omega}}^2 + \{\bI^{-1}(\bI \omega\times \omega)
%    		-\bI^{-1} \mu  \}^{\hat{}})\e \quad \hbox{Quadrotor equation}
%\end{equation*}
%\end{framed}
%
\begin{itemize}
  \item {\it Pendulum stabilization}: In order to apply a backstepping technique to this system of equations we first choose a desired vector $z_d(y(t), \dot{y}(t)) \in S^2$ which acts as the feedback control to the
  {\it pendulum equation} so that $y(t)$ is asymptotically stable about $\e$. To do so we define an
  intermediate control variable $f_p$ as follows:
	\begin{align*}
   		f_p \deff k_p  \e + k_d \dot{y}, \quad z_d(y, \dot{y}) = \frac{f_p}{ \norm{f_p}},
	\end{align*}
 and set the control $f$ as
 	\begin{align*}
		f \deff -Ml \norm{f_{p}}
	\end{align*}
   which renders
    \begin{equation*}
      \ddot{y} +\norm{\dot{y}}^2y = - {(\hat{y})}^2 f_p = - {(\hat{y})}^2[k_p  \e + k_d \dot{y}]
\end{equation*}
  \item  {\it Quadrotor stabilization}: In the next step, we wish to make the {\it quadrotor equation} track the trajectory $z_d(t)$.
   The choice of $z_d(y(t), \dot{y}(t))$ is to be made so that
   \begin{enumerate}
     \item The feedback control for stabilization of an SMS (given by \eqref{ureqd} in Lemma \ref{lem1}) is introduced through for stabilization of $y(t)$ at $\e$.
     \item $z_d(\e,0) =\e$ so that $z(t)$ is simultaneously stabilized about $\e$ along when $(y(t), \dot{y}(t)) = (\e,0)$.
   \end{enumerate}
  \item The choice of $\mu$ in \eqref{ddtz} is to be made such that the error variable $(z-z_d(t))$ is driven to zero in two steps. In the first step, the structure of an SMS on $S^2$ is imparted to \eqref{ddtz} and in the second the tracking control in Lemma \ref{lem2} is employed so that $z(t)$ asymptotically tracks the previously chosen $z_d(t)$.
  \item $k_p$ is chosen such that $\ddot{x}=0$ after the control objective is achieved.
\end{itemize}
%}
\subsection{Main Result}
In the following theorem, using appropriate Lyapunov functions we show  that the system of equations \eqref{ddty}-\eqref{ddtz} is asymptotically stable about $(y(t), \dot{y}(t), z(t), \dot{z}(t))= (\e,0,\e,0)$ for a suitable choice of $z_d(t)$, $\mu$ and $f$.

%The philosophy of the control design is to first choose $z= z_d$ so that $y$ asymptotically approaches $\e$ and then to choose the torque $\mu$ so that $z$ asymptotically approaches $z_d$.
%
\begin{theorem}

The following control thrust $f$ and moment $\mu$ ensures that $\lim_{t \to \infty} y(t) = \e$ and $\lim_{t \to \infty} z(t) = \e$
\begin{framed}
\begin{equation}\label{setf}
f \coloneqq -Ml \norm{f_p} \quad \text{and,}
\end{equation}
    \begin{align}\label{setmu}
    \{\bI^{-1}\mu\}^{\hat{}} \e \coloneqq & ( {\hat{\omega}}^2 + \{\bI^{-1}(\bI \omega\times \omega) \}^{\hat{}})\e\nn\\
     &+ R^\top(\norm{\dot{z}}^2 z - u_{fb}),
\end{align}
\end{framed}
where,
\begin{align}\label{fpeqn}
f_p \deff k_p  \e + k_d \dot{y},
\end{align}
 \begin{equation}\label{kpeqn}
  k_p= \frac{(M+m)g}{Ml (\norm{\e}+\norm{\dot{y}})},
\end{equation}
\begin{equation}\label{ufb}
    u_{fb} = -k_1 \hatsq{z} z_d - k_2 v_e + \nabla_{\dot{z}} (\tau(z,z_d) \dot{z}_d) - \hatsq{z} \beta,
\end{equation}
\begin{alignat*}{3}
   &\tau(z,z_d) \dot{z}_d \deff (z_d \times \dot{z}_d) \times z \quad
    &v_e \deff \dot{z} - \tau(z,z_d) \dot{z}_d,
\end{alignat*}
\begin{equation}\label{zdes}
  z_d(y, \dot{y}) = \frac{f_p}{ \norm{f_p}},
\end{equation}
$ \quad k_1, k_2, k_d \; \text{are positive constants, and,} \; \beta(t) \in \bR^3\; \text{ is defined as}$
 \begin{equation}\label{weqn}
      \beta= -\norm{f_p} A^\top (z-z_d)
 \end{equation}
 and is the minimum norm solution to $ A v_e = \dot{y},
      \text{where}
      A:T_z S^2 \to T_y S^2$.
\end{theorem}
\proof
Denote the desired equilibrium of \eqref{ddty} $(\e,0)=:(y^*,0)$. On substituting $f$ from \eqref{setf} and $z=z_d$ from \eqref{zdes} in \ref{ddty}, we obtain
\begin{equation}\label{ddtynew}
    \ddot{y} + \norm{\dot{y}}^2 y  = -k_p {(\hat{y})}^2 \e - k_d \dot{y}
\end{equation}
as ${(\hat{y})}^2 \dot{y} = \dot{y}$.  From Lemma \ref{lem1} it is observed that \ref{ddtynew} is asymptotically stable about $(y^*,0)$ for all $t$ such that $ z(t)= z_d(t)$. Subtracting $\frac{f}{Ml} \hatsq{y} z_d$ to both sides of \eqref{ddty} yields
\begin{equation}\label{ddtymod}
    \nabla_{\dot{y}} \dot{y} =  -k_p {(\hat{y})}^2 \e - k_d \dot{y} +\frac{f}{Ml} \hatsq{y} (z-z_d)
\end{equation}

Consider the Lyapunov function for the pendulum equation \eqref{ddtymod}
$ V_1 = k_p(1-y^\top \e)+ \frac{1}{2} \norm{\dot{y}}^2
$.
Therefore,
\begin{align*}
    \ddt{V_1} =& \ip{k_p {(\hat{y})}^2 \e}{\dot{y}} + \ip{\nabla_{\dot{y}} \dot{y}}{ \dot{y}} \\
    =& \ip{k_p {(\hat{y})}^2 \e}{\dot{y}} +  \ip{\frac{f}{Ml} \hatsq{y} (z-z_d)}{\dot{y}} \\
    &- \ip{k_p {(\hat{y})}^2 \e}{\dot{y}} - \ip{k_d \dot{y}}{\dot{y}} \\
    &= - \ip{k_d \dot{y}}{\dot{y}} + \ip{\frac{f}{Ml} \hatsq{y} (z-z_d)}{\dot{y}}
\end{align*}
%
%\textcolor{magenta}{Please conclude the above argument properly. There is a disconnect here.}
Next we look at the quadrotor subsystem in \eqref{ddtz}. The desired equilibrium is $z_d^*=\e$ and by choice, $z_d(y^*,0)= z_d^*$. Therefore both the dynamical equations \eqref{ddtz}-\eqref{ddty} attain their respective equilibria simultaneously. We choose $\mu$ so that $z(t)$ tracks $z_d(t)$. Substituting $\{\bI^{-1}\mu\}^{\hat{}} \e$ from \ref{setmu} in \ref{ddtz}
\begin{equation}\label{newddtz}
   \ddot{z}= -\norm{\dot{z}}^2 z + u_{fb} \; \; \text{which means},\;\; \nabla_{\dot{z}} \dot{z} = u_{fb}
\end{equation}
Since $z(t) \in S^2$, from Lemma \ref{lem2} we know that the feedback control that must be introduced to track $z_d(t)$ is $u_{fb}$ defined in \ref{ufb}. The Lyapunov function for the $z$ subsystem is chosen as:
\begin{align*}
V_2 = & k_1(1-z^\top z_d)+ \frac{1}{2} \norm{\dot{z} - \tau(z,z_d)\dot{z}_d}^2 \\
&= k_1(1-z^\top z_d)+ \frac{1}{2} \norm{v_e}^2
\end{align*}
The transport map $\tau(z,z_d): T_z S^2 \to T_{z_z}S^2$ is compatible with the potential function $k_1(1-z^\top z_d)$ (as defined in Theorem 11.19 in \cite{bulolewis}), therefore, $\ddt{k_1(1-z^\top z_d)} = \ip{k_1 {(\hat{z})}^2 z_d}{v_e}$ and,
\begin{align}\label{dv2}
    \ddt{V_2} &= \ip{k_1 {(\hat{z})}^2 z_d}{{v}_e} + \ip{\nabla_{\dot{z}} v_e} { v_e} \\ \nn
    &= \ip{k_1 {(\hat{z})}^2 z_d}{{v}_e} + \ip{\nabla_{\dot{z}} (\dot{z} - \tau(z,z_d)\dot{z}_d)}{ v_e}  \\ \nn
    &= \ip{k_1 {(\hat{z})}^2 z_d}{{v}_e} + \ip{\nabla_{\dot{z}} \dot{z} }{ v_e} -\ip{\nabla_{\dot{z}} \tau(z,z_d)\dot{z}_d}{ v_e}\\ \nn
    &= -\ip{k_2 v_e}{v_e} -\ip{\hatsq{z}\beta}{v_e}
\end{align}
Define a Lyapunov function $V \deff V_1 (y, \dot{y}) +V_2(z, \dot{z}, z_d, \dot{z}_d)$ for the entire system of equations \eqref{ddtynew}-\eqref{newddtz}, with $u_{fb}$ defined in \eqref{ufb}. By choosing $\beta$ according to \eqref{weqn} we ensure that the total Lyapunov function is nonincreasing, as follows.
\begin{equation*}
     \ddt{V} = \ddt{V_1+V_2}=  - \ip{k_d \dot{y}}{\dot{y}} -\ip{k_2 v_e}{v_e} \leq 0
\end{equation*}
\endproof

\begin{remark}\label{betarem}
In the expression for $\beta$ in \eqref{ufb}, $A:T_z S^2 \to T_y S^2$ is a transport map. $\beta$ is introduced as a feedback force to cancel the effect of the error between desired trajectory $z_d(t)$ and the state trajectory $z(t)$, which appears as $\frac{f}{Ml} \hatsq{y} (z-z_d)$ in \eqref{ddtymod}. Therefore $\beta(t)$ vanishes for all $t$ such that $z(t)= z_d(t)$.
\end{remark}
\subsection{Zero Dynamics}
In the above section, we use feedback control for the output $z(t)$ and only one state $y(t)$ of the system \eqref{xeqn_txt}-\eqref{ddty} by eliminating the other states from the dynamics of $y$ and $z$ variables. To understand the zero dynamics, we take a look at the effect of the applied control action on the other states $x(t)$ and $R(t)$. At the steady state, $(y(t), \dot{y}(t), z(t), \dot{z}(t))= (\e,0,\e,0)$ and $k_p= \frac{(M+m)g}{Ml}$ therefore, $v= k_p \e$, $\norm{v}=k_p$, $f= -Ml$ and the controlled trajectory $x(t)$ evolves as follows
\begin{equation}\label{xzero}
   \ddot{x}  = g\e -\frac{Ml k_p}{m+M} \e=g \e-g \e=0
\end{equation}
\section{Numerical Experiments}
%
%\textcolor{magenta}{Plots need to be redone. The timescale is not intuitive; please use seconds.
%Why not a quad with a pendulum schematic diagram, frozen at various instants of the stabilization
%manoeuvre  ? }
%
A quadrotor is considered with
\[\bI = diag([0.0820; 0.0845; 0.1377])kgm^2 \text{ and } M= 0.4kg.\]
The pendulum has
\[m =0.1m \text{ and }l=0.5m.\]
The initial conditions for orientation and position of the quad are
\begin{align*}
&R(0)= \begin{pmatrix}
0.36 &  0.48 & -0.8 \\
-0.8 & 0.6 & 0 \\
 0.48 & 0.64 & 0.60
\end{pmatrix}, \quad \omega(0)= \begin{pmatrix}
                                  0.8 \\
                                  -0.3\\
                                  0.5
                                \end{pmatrix},\\
&x(0) = \begin{pmatrix}
                                    1 \\
                                    1 \\
                                    1
                                  \end{pmatrix}, \quad \dot{x}(0)=\begin{pmatrix}
                                               2 \\1.5 \\1
                                             \end{pmatrix}
\end{align*}
The controller gains in \eqref{ufb} and \eqref{zdes} are denoted by $K= \begin{pmatrix}
             k_d & k_1 & k_2
           \end{pmatrix}$. The $\beta$ in \eqref{ufb} is computed using ``\textit{lsqminnorm}'' routine in MATLAB for all the experiments. As both $z(t)$ and $y(t)$ are in $S^2$, we plot the last coordinates which is shown to approach $1$ asymptotically thereby showing that both $z(t)$ and $y(t)$ approach $\e$ asymptotically. All the experiments are performed for the time $t \in \left[0s,6.5s \right]$. The initial bob positions, velocities and gain matrices $K$ are varied as:
          \begin{table}[h]
\resizebox{0.7\textwidth}{!}{\begin{minipage}{\textwidth}
\begin{tabular}{|c|c|c|c|c|}
  \hline
  Experiment No. &$y(0)$ & $\dot{y}(0)$& $K$ & Figure Number\\
  \hline
  1&$ \begin{pmatrix}
         1/\sqrt(2) \\
         0 \\
         1/\sqrt(2)
       \end{pmatrix}$ & $\begin{pmatrix}
                                    0.5 \\
                                    0 \\
                                    -0.5
                        \end{pmatrix}$  & $\begin{pmatrix}
                                           1 \\ 8 \\ 4
                                         \end{pmatrix}$ &Figure \ref{fig1}\\
                                         \hline
   2&$\begin{pmatrix}
         1/\sqrt(2) \\
         0 \\
         1/\sqrt(2)
       \end{pmatrix}$ & $\begin{pmatrix}
                                    0.7 \\
                                    0 \\
                                    0.7
                        \end{pmatrix}$ & $\begin{pmatrix}
                                           1 \\ 9 \\ 4.4
                                         \end{pmatrix}$ & Figure \ref{fig2}\\
  \hline
  3&$\begin{pmatrix}
         0.1 \\
         0.0995 \\
        -0.99
       \end{pmatrix}$ & $\begin{pmatrix}
                                    2.2263 \\
                                    0.25 \\
                                    0.25
                                  \end{pmatrix}$& $\begin{pmatrix}
                                                    1 \\
                                                    11 \\
                                                    5
                                                  \end{pmatrix}$&Figure \ref{fig3}\\
  \hline
  4&$\begin{pmatrix}
         0 \\
         0\\
        -1
       \end{pmatrix}$ & $\begin{pmatrix}
                                    0 \\
                                    0 \\
                                    0
                                  \end{pmatrix}$& $\begin{pmatrix}
                                                    1 \\
                                                    12 \\
                                                    5
                                                  \end{pmatrix}$&Figure \ref{fig4}\\
  \hline
  5&$\begin{pmatrix}
         -1/\sqrt(2) \\
         0 \\
         1/\sqrt(2)
       \end{pmatrix}$ & $\begin{pmatrix}
                                    0.7 \\
                                    0 \\
                                    0.7
                                  \end{pmatrix}$& $\begin{pmatrix}
                                                    1 \\9 \\4
                                                  \end{pmatrix}$&Figure \ref{fig5}\\
  \hline
\end{tabular}
\end{minipage} }
\end{table}

%
%\textcolor{magenta}{It would be nice to see the other axes as well. The timescale on this
%plot needs to change.}
%

\begin{figure}[h!]
\begin{multicols}{2}
    \includegraphics[width=\linewidth]{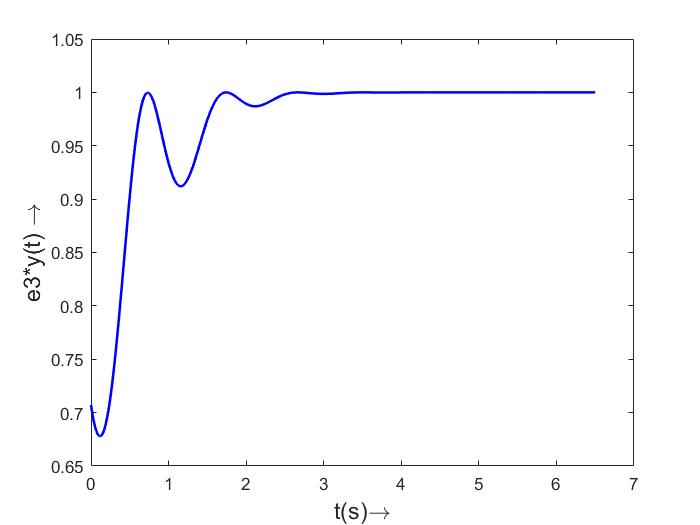}\par
    \includegraphics[width=\linewidth]{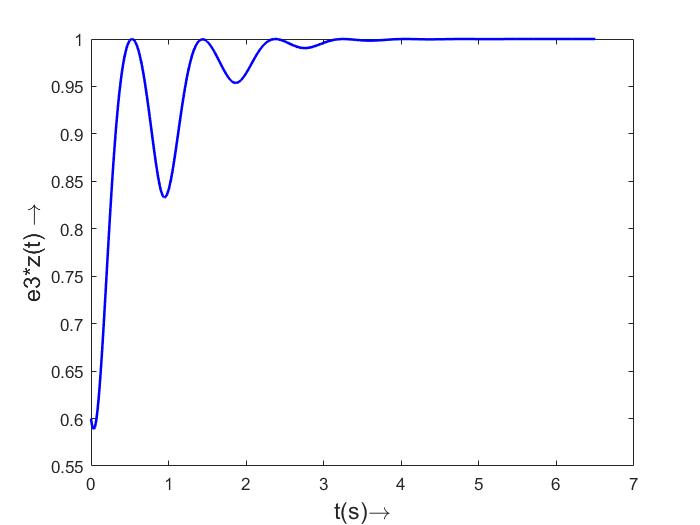}\par
    \end{multicols}
\caption{Experiment 1}
\label{fig1}
\end{figure}

\begin{figure}[h!]
\begin{multicols}{2}
    \includegraphics[width=\linewidth]{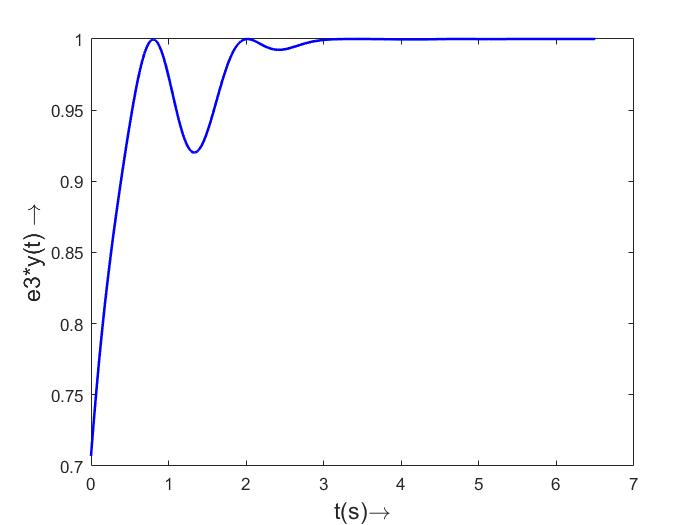}\par
    \includegraphics[width=\linewidth]{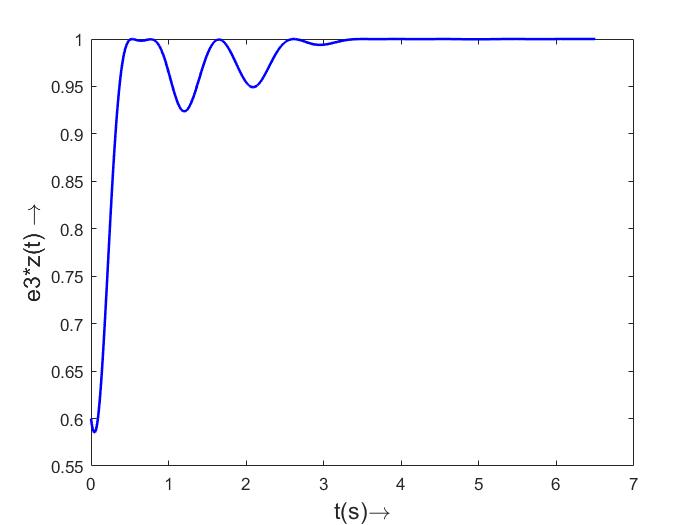}\par
    \end{multicols}
\caption{Experiment 2}
\label{fig2}
\end{figure}

\begin{figure}[h!]
\begin{multicols}{2}
    \includegraphics[width=\linewidth]{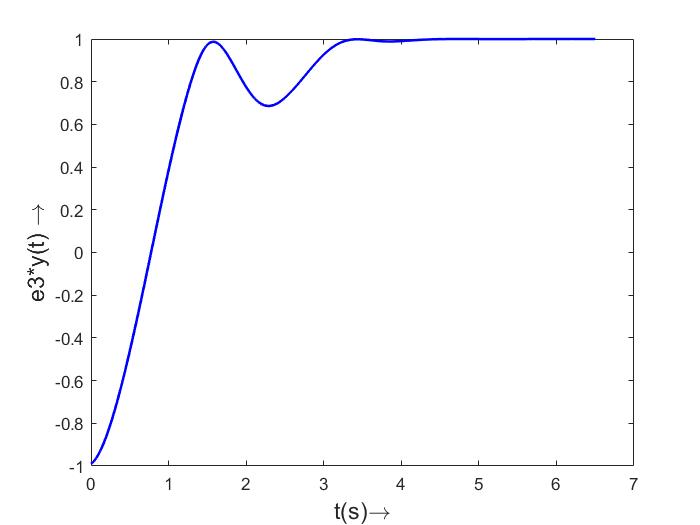}\par
    \includegraphics[width=\linewidth]{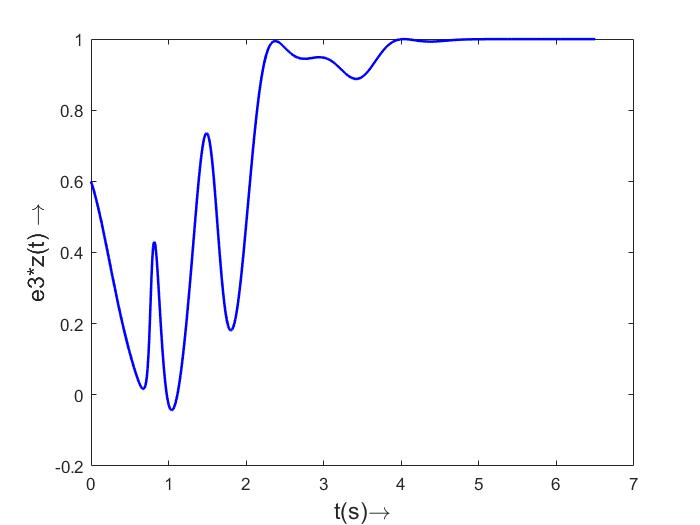}\par
    \end{multicols}
\caption{Experiment 3}
\label{fig3}
\end{figure}
\begin{figure}[h!]
\begin{multicols}{2}
    \includegraphics[width=\linewidth]{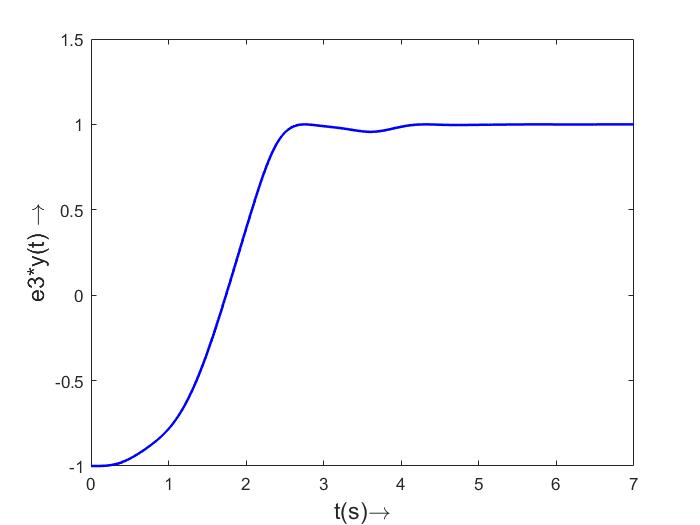}\par
    \includegraphics[width=\linewidth]{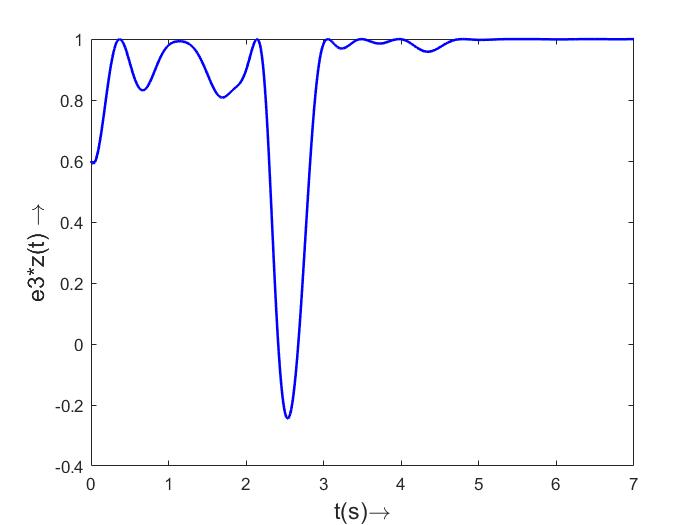}\par
    \end{multicols}
\caption{Experiment 4}
\label{fig4}
\end{figure}
\begin{figure}[h!]
\begin{multicols}{2}
    \includegraphics[width=\linewidth]{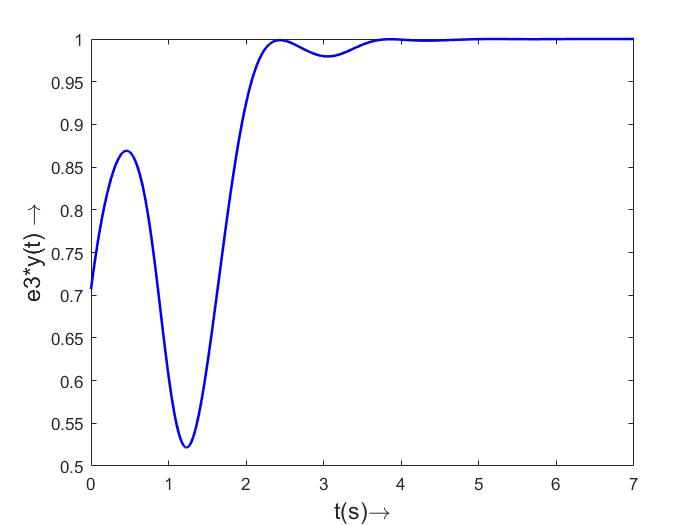}\par
    \includegraphics[width=\linewidth]{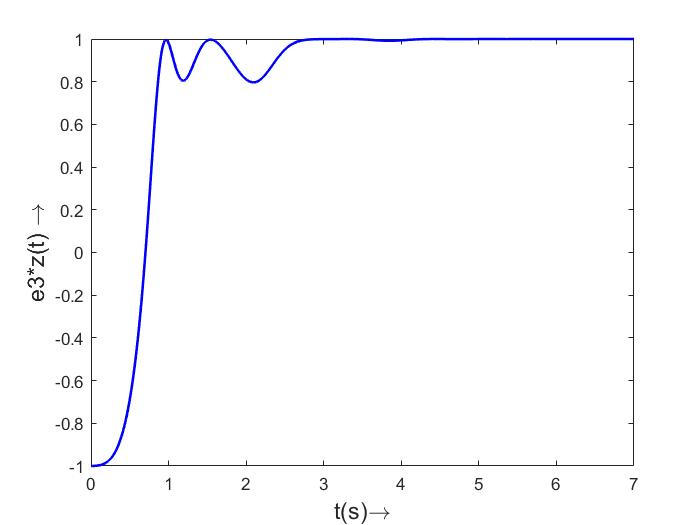}\par
    \end{multicols}
\caption{Experiment 5}
\label{fig5}
\end{figure}
\begin{figure}
  \centering
  \includegraphics[width=\linewidth]{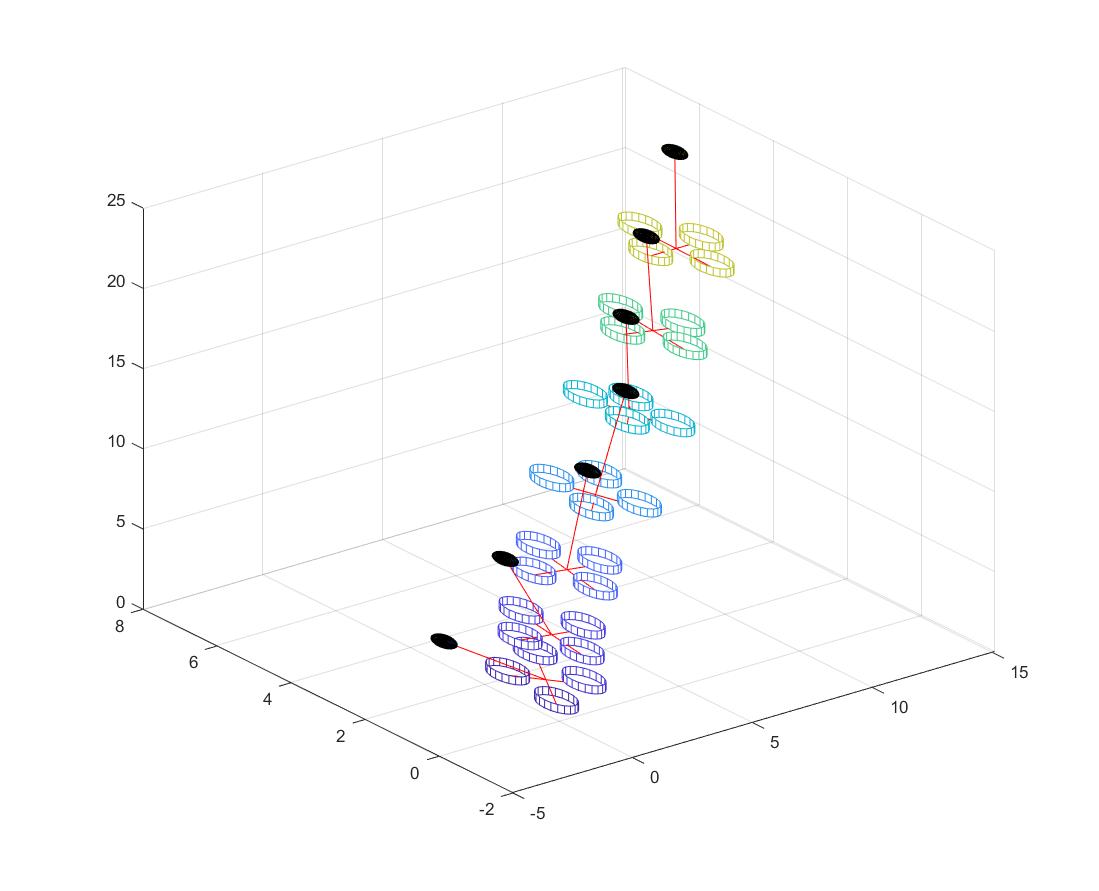}
  \caption{Stick figure in stop motion for Experiment 2}\label{stq1}
\end{figure}
\begin{figure}
  \centering
  \includegraphics[width=\linewidth]{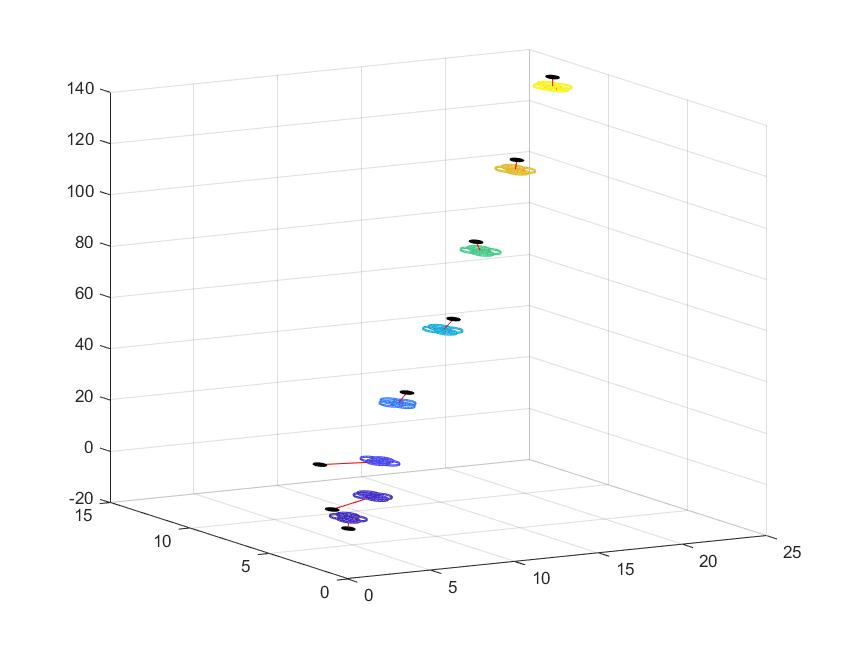}
  \caption{Stick figure in stop motion for Experiment 4}\label{stq2}
\end{figure}

The Figures \ref{stq1} and \ref{stq2} show the pivot and the bob of the pendulum in $S$ frame in stop motion for experiments 2 and 4. The following: \url{https://www.youtube.com/watch?v=AZQobVlMb1U} and \url{https://www.youtube.com/watch?v=_R1zeMYo0aE} show slow motion movie in MATLAB for Experiments 4 and 5 respectively for a camera mounted on $-\e$ axis.
%\begin{figure}
%  \centering
%  \includegraphics[width=\linewidth]{quad_pendulum_timeseries_5.jpg}
%  \caption{Stick figure in stop motion for Experiment 5}\label{stq3}
%\end{figure}

%$t(10^{-3} s) \rightarrow$
%$e3*z(t) \rightarrow$

\section{Conclusion}
In this paper, the problem of swing up of a spherical pendulum mounted on a quadrotor was studied assuming that the quad is actuated by free wheel rotors and thus can actuate the pendulum. A feedback control was proposed based on geometric principles leading to stabilization of the inverted payload (bob of the pendulum) from a large set of initial conditions. The quad was also stabilized so that the yaw orientation aligned with the upright pendulum position. The results presented were verified with the help of numerous MATLAB experiments. The results in this paper can be therefore applied to balance transport mounted on a UAV. The effect of disturbances needs to be carefully examined and is thus a challenging future direction as exponential stability cannot be guaranteed with the proposed control in this paper.

\bibliographystyle{plain}
\bibliography{aps1}
\end{document}